\documentstyle[preprint,aps,epsfig]{revtex}  

\tighten       

\begin{document}

\draft
\preprint{Version of \today }

\title{On the Photorefractive Gunn Effect}

\author{Luis L. Bonilla \and Manuel Kindelan
 \and Pedro J. Hernando }

\address{Universidad Carlos III de Madrid, Escuela Polit{\'e}cnica 
Superior, 28911 Legan{\'e}s, Spain}


\maketitle

\begin{abstract}
We present and numerically solve a model of the photorefractive Gunn effect. 
We find that high field domains can be triggered by phase-locked interference 
fringes, as it has been recently predicted on the basis of linear 
stability considerations. Since the Gunn effect is intrinsically nonlinear, 
we find that such considerations give at best order-of-magnitude estimations
of the parameters critical to the photorefractive Gunn effect. 
The response of the system is much more complex including multiple 
wave shedding from the injecting contact, wave suppression and chaos 
with spatial structure.
\end{abstract}

\narrowtext

\section{Introduction}
Segev, Collings and Abraham (SCA)~\cite{segev96} proposed an 
interesting mechanism for producing Gunn domains by means of a photorefractive 
parametric excitation. It occurs when two optical waves of slightly 
different frequencies are incident upon a biased
semiconductor crystal doped with deep impurity centers.
The authors conjecture that
the resulting traveling interference pattern excites
multiple high-field Gunn domains which move phase locked
with the interference fringes. Recently, L. Subacius,
{\em et al.}, \cite{subacius97} proposed an efficient way
of creating simultaneously a number of quasilocalized
high field Gunn domains through hot carrier transport in
spatially modulated and nonuniformly heated electron-hole
plasma, and present some numerical results and preliminary
experimental confirmation.

In this paper we present a consistent model of the
photorefractive Gunn effect and carry out numerical
simulations to understand the dynamics of the system.  
We find that high field domains 
can indeed be triggered by phase-locked interference fringes, as suggested
by SCA. However, the response of the system can be very
complex, and it is not possible to use a simplified
version of Kroemer's NL criterion, as suggested by SCA,
to predict the number of high field Gunn domains 
traveling through the sample. Indeed, our results 
indicate that with appropriate values of the parameters
of the system the response becomes chaotic
and this is, therefore, another example 
of driven (sinusoidal interference pattern of intensity $I$) chaos.

\section{Model equations}
The following equations describe the 
photorefractive Gunn effect,
\begin{eqnarray}
{\partial N_D^i\over\partial t} \, &=& \, S I (N_D \, - \,  N_D^i) \, -
\, \gamma n N_D^i, 
\label{d1} \\
{\partial n \over \partial t }\, - {\partial N_D^i\over\partial t} \, &=& \,
{1\over q} \, {\partial J \over \partial z},
\label{d2} \\
{\partial E \over \partial z} \, &=& \, - {q\over\varepsilon_s}\, (n \, + 
\, N_A \, -  \, N_D^i)
\label{d3} 
\end{eqnarray}
where (\ref{d1}-\ref{d2}) are the continuity equations for ionized
donors and for electrons respectively, and  (\ref{d3}) is Poisson's law.
In these equations, $z$, is the space variable in the direction
of current flow, $t$, is the time,
$N_D^i(z,t)$, represents the number density of ionized
donors,  $N_D$, the total donor density, 
$I(z,t)$, the incident light intensity, 
$S$, the photoionization
cross section, $\gamma$, the recombination rate, $n(z,t)$, the electron
number density, $-q$ the electron charge,  $J(z,t)$, the current density,
$E(z,t)$, the space-charge field inside the crystal, 
$N_A$, the density of negatively charged acceptors,
and,
$\varepsilon_s$, the low frequency dielectric constant.

The light intensity, $I(z,t)$, is given by
\begin{eqnarray}
I(z,t) \, = \, I_0 \, \left[1 \, + \, m \, \cos (Kz \, + \, \Omega t)
\right]
\label{intens} 
\end{eqnarray}
which, as decribed in  \cite{segev96}, is the intensity
of the moving
interference pattern formed  when two quasimonochromatic plane waves
of slightly different frequencies, $\omega$ and $\omega + \Omega$
($\Omega << \omega$) and slightly different angles of incidence
illuminate a bulk semiconductor crystal. In (\ref{intens})
$K=2 \pi / \Lambda$ is the interference wave number,
$m$ the modulation depth of the interference grating, and
$I_0$ the total average intensity.

Following Sze~\cite{sze81} [Eq.~(28) in chapter 11]
the current density, $J$, includes drift and diffusion terms,
\begin{eqnarray}
J \, = \, q n v(E) \, + q \, \frac{\partial \left[D(E)
n\right]}{\partial z}.
\label{j1}
\end{eqnarray}
which is the standard form of the drift-diffusion current 
density \cite{shaw79,HB92}
where, $v(E)$, is the electron drift velocity, and $D(E)$ the diffusion
coefficient.
The drift velocity of the electrons is a known function of the electric
field exhibiting negative differential resistance. In the following
analysis we use a saturating drift velocity function given by,
\begin{eqnarray}
v(E) \, = \, v_s \, \left[
1 \, + \, \frac{E/E_s \, - \, 1}{1 \, + \, A (E/E_s)^{\beta}} \right].
\label{vdee}
\end{eqnarray}
where $v_s$ is the saturation drift velocity, $E_s$ the saturation
field,
and $A$ and $\beta$ dimensionless constants that depend explicitly on
the mobility of the material ($\mu = v_s/E_s$). 

Differentiating the Poisson equation (\ref{d3}) with respect to time,  
inserting the result in the  continuity equation for
electrons  (\ref{d2}), and  
integrating with respect to space, results in
$\varepsilon_s \partial E/\partial t + J = J_{total}$. Here the constant
of integration, $J_{total}$, is the total current density. Introducing the electron current density 
given by (\ref{j1}) in the previous equation,
leads to  Amp{\`e}re's equation,
\begin{eqnarray}
J_{total} \, = \, q n v(E) \, + q \, \frac{\partial \left[D(E)
n\right]}{\partial z} \, + \,
\varepsilon_s \, \frac{\partial E}{\partial t}.
\label{ampere}
\end{eqnarray}
Notice that in Eq.~(6) of \cite{segev96} the current
density erroneously 
includes the displacement term, and this
leads to an incorrect Amp{\`e}re's equation
in which the displacement current drops out.

In addition to these equations the electric field distribution
must satisfy the {\em reverse bias} condition for a given applied 
voltage $V$,
\begin{eqnarray}
\int_0^L \, E \, dz \, = \, V.
\label{bias}
\end{eqnarray} 
Thus, to model the  photorefractive Gunn effect
we use the correct form of 
Amp{\`e}re's law (\ref{ampere}), the continuity equation
for ionized donors (\ref{d1}), Poisson's law (\ref{d3}),
the bias condition (\ref{bias}), and 
appropriate 
initial and boundary conditions. Solution of these
equations using the known functions for the
light intensity (\ref{intens}) and drift velocity
(\ref{vdee}),
provide the four unknowns, $N_D^i(z,t)$, 
$n(z,t)$, $E(z,t)$ and $J_{total}(t)$. 

It should be emphasized that boundary conditions
play an essential role in the existence of the Gunn
effect. Although  it can be debated which are
the correct conditions to apply in order to
simulate a particular experiment, 
periodic boundary conditions, as used in \cite{segev96}, 
should be avoided: they yield 
a total current density which is 
constant in time when a dc voltage bias is imposed.  
In fact,
integrating Amp{\`e}re's law we find $J_{total} = L^{-1} [\varepsilon_s
dV/dt + \int_0^{L} J dx]$. Then imposing periodic boundary conditions and a 
dc voltage bias we find that $dJ_{total}/dt = (c/L) \int_0^{L} 
(\partial J/\partial x) dx = 0$ (provided we have a domain moving at constant 
speed $c$; note that with physically reasonable boundary conditions, the domain 
cannot move at constant speed as it arrives at a realistic contact, e.g.\ 
Ohmic). Since the Gunn effect refers to time-dependent
oscillations of the current under dc voltage bias, one should not
use periodic boundary conditions when discussing it.

The most important boundary condition in a study of the Gunn effect 
in long samples is that for the injecting contact. In fact, the formation 
of Gunn domains is due to a periodic destabilization of the boundary
layer attached to such contact during the oscillations \cite{shaw79,HB92}. 
During the formation of a new wave at the 
injecting boundary the displacement current plays
a crucial role and can not be neglected.
Thus, we use an ohmic condition at the injecting 
contact,
$E = \rho \, J_{total}$ 
\cite{HB92,slowg,llbpre,inmapre}, 
and $\partial E/\partial z = 0$ at the receiving contact (which is 
thus passive and integration time is saved). 
When $\rho$ is such that $E/(\rho q N_A)$ intersects 
$v(E)$ on its decreasing branch, the Gunn effect is found for $m=0$ and 
appropriate values of the bias $V$ \cite{shaw79,HB92}. 

In order to solve numerically our model equations, we will first write them 
in nondimensional form. Let us redefine our variables in dimensionless form 
as follows: 
\begin{eqnarray}
& \begin{array}{lclclclclcl}
y \, &=& \, \displaystyle \frac{1}{L} \, z, & \qquad & 
s  \, &=& \, \displaystyle \frac{v_s}{L} \, t,  & \qquad &
F  \, &=& \, \displaystyle \frac{E}{E_s}, 
\end{array} \nonumber \\ \\
& \begin{array}{lclclclclclclcl}
j \, &=& \, \displaystyle \frac{J_{total}}{v_s q N_A}, & \qquad   &
p  \, &=& \, \displaystyle \frac{N_D^i}{N_A}, & \qquad  & 
\eta \, &=& \, \displaystyle  \frac{n}{N_A}, & \qquad  & 
\nu (F) \, &=& \, \displaystyle \frac{v(F E_s)}{v_s}\,.
\end{array} \nonumber
\end{eqnarray}
Inserting this in (1), (3)-(8), we obtain 
\begin{eqnarray}
& & \epsilon \, \left( {\partial F\over\partial s} \, 
+ \, \delta \, {\partial \eta\over\partial y}
\right) \, = \, j(t) \, - \, \eta \, \nu(F) ,
\label{nd1} \\
& & \epsilon \, {\partial F\over\partial y} \, = \, p \, - \, 1 \, - \, \eta ,
\label{nd2} \\
& & {\partial p\over\partial s}\, = \, i_0 \, \left[1 \, + \, m\,
\cos{(ky \, + \, w s)} \right] \, (1 \, - \, 
 \alpha \, p)
\, - \, \beta \, \eta \, p ,
\label{nd3} \\
& & \int_0^1 F \, dy \, = \, \phi.
\label{nd4}
\end{eqnarray}
Here we have defined the following nondimensional parameters (we assume 
constant diffusivity, $D(F)=D$):
\begin{eqnarray}
& \begin{array}{lclclclclclclclclcl}
\epsilon \, &=& \, \displaystyle \frac{\epsilon_s E_s}{q N_A L},&\quad &
\phi \, &=& \, \displaystyle \frac{1}{E_s L} \, V, &\quad &
\delta \, &=& \, \displaystyle \frac{q D N_A}{v_s \epsilon_s E_s}, &\quad &
\beta \, &=& \, \displaystyle \frac{\epsilon_s E_s}{\epsilon v_s q} \, \gamma 
\, ,
\end{array} \nonumber \\ \\
& \begin{array}{lclclclclclclcl}
\alpha \, &=& \, \displaystyle \frac{N_A}{N_D},& \qquad&
i_0 \, &=& \, \displaystyle \frac{S \epsilon_s E_s N_D}{\epsilon v_s q N_A^2} \, I_0, &\qquad&
w \, &=& \, \displaystyle  \frac{\epsilon_s E_s}{\epsilon v_s q N_A} \, \Omega, &\qquad &
k \, &=& \, \displaystyle \frac{\epsilon_s E_s}{\epsilon q N_A} \, K .
\end{array} \nonumber
\end{eqnarray}

\section{Results}
We have solved (\ref{nd1}) - (\ref{nd4}) numerically for different values of 
the parameters $w$ and $i_0$, keeping fixed $\epsilon = 4.138\, 10^{-4}$, 
$m = 0.1$, $\rho = 1.8$, $k = 2\pi$, $\delta = 0.05$, $\alpha = 0.01$, 
$\beta = 3.5294$, and $\phi=3$. These numerical values correspond to those 
used by SCA~\cite{segev96}. We were particularly interested in testing 
a central piece of SCA's analysis, namely their particular use of Kroemer's 
NL criterion, Eq.~(20). Prompted by SCA's suggestion that the length in 
Kroemer's criterion may be the distance between multiple domains, we combine 
Eq.~(20) with Eq.~(17) for the electron density $n_1$, and use $\ell=L/N$,
where $\ell$ is the distance between adjacent domains.
Then we obtain a formula for the maximum 
number ($N$) of high field domains which may coexist:
\begin{equation}
N < N_0 \equiv {195.565\over 1 + {10^{11}\over I_{0}} - 100 \left({\Omega\over 
I_{0}} \right)^{2}}\,. \label{1}
\end{equation}
We have used SCA's numerical values ($I_0$ and $\Omega$ are 
measured in W.m$^{-2}$ and Hz, respectively). Fig.~1 
depicts the level curves $\Omega(I_0)$ which are obtained 
when $N_0$ takes on different integer values. If SCA's
theory holds, $N = i$ between the level curves $N_0=i-1$ and $N_0=i$. However,
we observed Gunn domains (Fig.~2--A) where Segev et al's
theory predicts they should not be (point A in Fig.~1). Points
C and D in Fig.~1 would correspond to coexistence of $N=2$ and $N=3$ domains respectively, but we
find four domains 
and a periodic response (Fig.~2--C) and a quasi chaotic
response  (Fig.~2--D) respectively. Finally,  
at point B corresponding to the
conditions proposed by SCA to illustrate their theory, 
one domain is created during each oscillation 
period (Fig.~2--B). 
In fact, the response of the system is very complex
and within each of those zones in the $\Omega(I_0)$
plane, where a constant number of high field domains
is predicted it is possible to find all kinds of
behavior: periodic with a single frequency, periodic
with a high number of frequencies and chaotic.

The discrepancy between the computed and
predicted response should not be surprising. 
In fact,
the Gunn effect is a periodic oscillation of the current 
($\propto J_{total}$) in a dc voltage biased semiconductor 
presenting negative differential velocity. It is due to periodic shedding 
and motion of charge dipole waves (high field domains) at a boundary or 
a nucleation site. It is intrinsically nonlinear 
\cite{sze81,shaw79,HB92,slowg,llbpre,inmapre} 
so that the relevance of
linear approximations such as those used
in \cite{segev96} is questionable.
We think that  Kroemer's NL criterion implies that
for a given model (equations, bias, boundary and 
initial conditions) it can be proved that no oscillatory instability 
is possible unless the NL product (where L is semiconductor length and 
N is doping density) is above a certain number. For Kroemer's
model of the Gunn effect in n-GaAs
under dc voltage bias see Figure 3
of \cite{siam94}.
An analytical estimate (probably not a very precise one) could be obtained by
adapting to the Kroemer model 
the arguments in the Appendix of 
\cite{wacker}. This said, SCA have used a particular 
version of the NL criterion in which N is an electron density and L 
is either the semiconductor length (one domain) or the distance between
domains (multiple domains). It is not surprising 
that this usage produces results that are
not quantitatively correct.

It is interesting to compare the response of the system 
in points C and D of Figure 1. We observe that the dimensionless
current density, $j(t)$, in case C is periodic (Fig.~2)
although it contains a large number of frequencies,
as can be seen in the FFT spectrum shown in the
lower right of the Figure. Instead, the response in case
D is not periodic and the FFT spectrum appears chaotic.
The difference between these two cases can be understood
by looking at the electric field profiles as a function of time. 
In case C, there are several waves that originate
at the injecting contact (right contact) and propagate
towards the receiving contact. Each of these waves
does not overtake the previous one and this leads to a
smooth, periodic response. In case D, however, a wave
originates at the injecting contact and propaggates
through the sample. Later, a new wave is injected which
moves faster than the previous one, and that overtakes
it before reaching the receiving contact. The interaction
between waves inside the sample, 
leads to a complex response which may become chaotic.

To understand the transition between the periodic and
chaotic solutions we carried out numerical simulations
for a fixed value of $\Omega$ and varying the intensity
$I_0$ within a certain small range corresponding to
line E in Figure 1. The results of these simulations
are summarized in Figure 3
which shows the intensity of each mode
of the FFT spectrum as a function of $I_0$.
Notice that there is a frequency, which is very close to the
excitation frequency, $\Omega$, which appears for all values of
$I_0$. Also, all integer multiples of this frequency are present
in the current density history.   
It can also be observed that there is a sharp transition between
a periodic solution and a chaotic one. Figure 4 shows
the current density history and field profiles for case
F in the periodic region, and for case G inside the
chaotic region. 

The bias condition imposes a conservation of area in the electric
field profile $E(z)$ for all time. Thus, the size of a travelling wave 
must decrease when a new wave is generated in the injecting contact 
(right contact).
If the new wave grows fast enough the old wave must disappear to
keep the area constant.
In both cases, see F and G in Fig. 4, several waves are born
at the injecting contact and propagate towards the receiving contact
during a period.
In case F, all the waves reach the left contact and this leads to a
periodic response in the density current.
As the intensity $I_0$ increases, the wave formation velocity increases
until a critical value above which the waves that propagate through the
sample disappear before reaching the receiving contact.  
This produces a complex behavior with different patterns of 
formation-disappearance of waves.
Larger values of $I_0$ result in chaotic responses such as that shown in  
case G as an example.

It is also interesting to observe that within the
chaotic region there are windows in which the response
becomes again periodic, with frecuency given by the fundamental frequency
divided by 2, 3 or 4.
In these specific ranges of $I_0$, the interaction
between waves inside the sample produces a simple pattern 
with a periodic sequence.

\section{Conclusions}
We have thus found that high field domains 
could indeed be triggered by phase-locked interference fringes, as suggested
by SCA. However a literal use of their version of Kroemer's NL criterion
does not often agree with numerical simulations of the model for photorefractive 
Gunn effect.  In addition, we have found very interesting examples of natural 
and driven (sinusoidal interference pattern of intensity $I$) chaos
when appropriate parameter values are used. In the regime for which
the high field domains carry a large fraction of the bias, an asymptotic theory 
of the Gunn effect \cite{HB92,slowg,llbpre,inmapre} can be extended 
and applied to the photorefractive model, and used to
interpret and predict the results of the numerical simulations with
greater accuracy than SCA's linearized approach.
These results will be presented elsewhere. 


\newpage
\begin{figure}
\caption{Level curves of intensity versus driving frequency for 
different integer values of $N_0$ in Eq.~(1). Numerical simulations
have been performed for points labelled A, B, C and D, with dimensionless
values of intensity and frecuency $(i_0,w)$: A=(0.0025,0.05), B=(0.035,0.055), 
C=(0.05,0.0109), D=(0.07,0.05). The Fourier spectrum of the current will 
be displayed later in Figure 3 for the points on the horizontal line E.
}
\label{fig1}
\end{figure}

\begin{figure}
\caption{Numerical simulations for values of the current and the oscillation 
frequency corresponding to points A, B, C and D in Figure 1. All other 
parameters correspond exactly to those proposed in~[1] (see text). Each Figure 
contains four graphics in dimensionless variables: a spatio-temporal density 
plot of the electric field, $F(y,s)$ (density scale on the upper right part), 
a plot of the current density $j(s)$ (upper left), the power spectra of the 
$j(s)$ signal in arbitrary units (lower right) and the electric field as a 
function of $y$ for the fixed time marked with a white horizontal line 
in the $F(y,s)$ density plot (lower left).  
}
\label{fig2}
\end{figure}

\begin{figure}
\caption{Fourier power spectra of $j(t)$  for $w=0.05$ and diferent values 
of $i_0$ in the range $0.045-0.065$, (line E of Figure 1). Each FFT is 
represented by a narrow horizontal band with gray scaled frecuency modes 
amplitudes: white (large) and black (small), in arbritary units. The arrows 
correspond to the simulations of a periodic signal (F) and a chaotic one (G) 
carried out in Figure 4.  
}
\label{fig3}
\end{figure}

\begin{figure}
\caption{Numerical simulations for the points F and G in Figure 3. The layout 
of the graphics is the same as in Figure 2. 
}
\label{fig4}
\end{figure}


\end{document}